\documentclass[useAMS,usenatbib]{mn2e}

\usepackage{graphicx}

\newcommand{\dfrac}[2]{{\displaystyle \frac{#1}{#2}}  }
\newcommand{\vect}[1]{\mbox{\boldmath$#1$}}
\newcommand{\ro  }{a_{\rm p}} 
\newcommand{\cm  }{\,{\rm cm}^{-3} } 
\newcommand{\gcm  }{\,{\rm g\,cm}^{-3} } 
\newcommand{\scm  }{\,{\rm g\,cm}^{-2} } 
\newcommand{\etal  }{{et al.} }
\newcommand{\msun}{\thinspace M_\odot}  
\newcommand{\hs}{h}
\newcommand{\rh  }{r_{\rm H}} 
\newcommand{\rht }{\tilde{r}_{\rm H}} 
\newcommand{\mj}{M_{\rm J}}
\newcommand{\me  }{\thinspace M_\oplus } 
\newcommand{\tl}[1]{\tilde{#1}}
\newcommand{\deft}{\tilde{t}}
\newcommand{\rhoc}{\rho_{\rm cri}}
\newcommand{\jcm}{{\rm cm^2\,s^{-1}}}
\newcommand{\rp  }{r_{\rm p}}

\title[Circumplanetary Disk Formation]
  {Thermal Effects of Circumplanetary Disk Formation around Proto-Gas Giant Planets}
\author[M. N. ~Machida]
  {M. N. ~Machida \\
  $^1$Department of Physics, Graduate School of Science, Kyoto University, Sakyo-ku, \\ 
Kyoto 606-8502, Japan; machidam@scphys.kyoto-u.ac.jp}
\pagerange{\pageref{firstpage}--\pageref{lastpage}} \pubyear{2008}

\def\LaTeX{L\kern-.36em\raise.3ex\hbox{a}\kern-.15em
    T\kern-.1667em\lower.7ex\hbox{E}\kern-.125emX}

\begin{document}

\label{firstpage}

\maketitle
\begin{abstract}
The formation of a circumplanetary disk and accretion of angular momentum onto a protoplanetary system are investigated using three-dimensional hydrodynamical simulations. 
The local region around a protoplanet in a protoplanetary disk is considered with sufficient spatial resolution: the region from outside the Hill sphere to the Jovian radius is covered by the nested-grid method.
To investigate the thermal effects of the circumplanetary disk, various equations of state are adopted.
Large thermal energy around the protoplanet slightly changes the structure of the circumplanetary disk.
Compared with a model adopting an isothermal equation of state, in a model with an adiabatic equation of state, the protoplanet's gas envelope extends farther, and a slightly thick disk appears near the protoplanet.
However, different equations of state do not affect the acquisition process of angular momentum for the protoplanetary system.
Thus, the specific angular momentum acquired by the system is fitted as a function only of the protoplanet's mass.
A large fraction of the total angular momentum contributes to the formation of the circumplanetary disk. 
The disk forms only in a compact region in very close proximity to the protoplanet.
Adapting the results to the solar system, the proto-Jupiter and Saturn have compact disks in the region of $r<21\,$$r_{\rm Jup}$ ($r<0.028$\,$r_{\rm H,Jup}$) and $r<66$\,$r_{\rm Sat}$ ($r<0.061$\,$r_{\rm H,Sat}$), respectively, where $r_{\rm Jup}$ ($r_{\rm H,Jup}$) and $r_{\rm Sat}$ (r$_{\rm H,Sat}$) are the Jovian and Saturnian (Hill) radius, respectively.
The surface density has a peak in these regions due to the balance between centrifugal force and gravity of the protoplanet.
The size of these disks corresponds well to the outermost orbit of regular satellites around Jupiter and Saturn.
Regular satellites may form in such compact disks around proto-gas giant planets.
\end{abstract}


\section{Introduction}
Since the first detection by \citet{mayor95}, over 300 extrasolar planets (or exoplanets) have been observed. 
Almost all exoplanets are believed to be gas giant planets, like Jupiter and Saturn in the solar system, because massive planets are observed preferentially. 
Although these planets are believed to form  in the disk surrounding the central star (i.e., the circumstellar disk or protoplanetary disk), their formation process is not yet fully understood.
Under spherical symmetry, including the radiative effect, the evolution of gas giant planets has been investigated \citep[e.g.,][]{mizuno78,mizuno80,stevenson82,bodenheimer86,pollack96,ikoma00,ikoma01}.
These studies showed rapid gas accretion onto the protoplanet after the formation of a solid core with $5-20\me$.
The gas giant planet acquires a large fraction of its mass in this gas accretion phase.
Since gas flows into the gravitational sphere (i.e., Hill sphere) of the protoplanet with an angular momentum, a circumplanetary disk forms around the protoplanet in this phase.
Then, satellites may form in this circumplanetary disk.
Regular satellites around a gas giant planet are considered to form according to a scenario similar to the formation of earth-like planets in the solar system: the protosatellite forms by the accumulation through mutual collision of the satellitesimals that forms after the dust grains sink towards the equatorial plane of the circumplanetary disk \citep[e.g.,][]{stevenson86}.
Since regular satellites are considered to be a by-product of gas-planet formation, they provide an important clue for understanding it.
Thus, it is necessary to understand the formation of the circumplanetary disk, which is the  site of the satellite formation in the framework of gas planet formation.
However, the formation and acquisition process of angular momentum in the circumplanetary disk cannot be understood by spherically symmetric calculations.

In two-dimensional simulations, protoplanet evolution was investigated by many authors  \citep[e.g.,][]{sekiya87, lubow99, kley99, dangelo02, tanigawa02,dobbs07} who showed that it is possible for a circumplanetary disk to form around a protoplanet.
They also found prograde rotation of the protoplanetary system (protoplanet + circumplanetary disk) against the orbital motion of the planet in the circumstellar disk.
Recently, the evolution of a protoplanetary system was calculated in three dimensions \citep[e.g.,][]{miyoshi99, kley01,bate03,dangelo03,machida08}.
These studies showed that the gas flow pattern in three dimensions is qualitatively different from that in two dimensions: gas flows into the protoplanet system only in the vertical direction in three-dimensional simulations.
Thus, three-dimensional simulations are necessary for investigating the gas flow around the protoplanet, angular momentum of the protoplanet system, and circumplanetary disk formation.
Moreover, the circumplanetary disk should be resolved with sufficient spatial resolution, because regular satellites are distributed only in very close proximity to the planet; the regular satellites of giant planets in the solar system are distributed only in the region of $r \le 60\rp$, where $\rp$ is the radius of the planet.
On the other hand, the radius of the gravitational sphere (or Hill radius) of the planet extends up to $\rh\sim 1000\,\rp$.
For example, the Hill radii for Jupiter and Saturn are $r_{\rm H,J}=743\,\rp$  and $r_{\rm H,S}=1083\,\rp$, respectively.
Thus, to investigate a protoplanetary system, we must resolve not only the planetary radius $\sim\rp$, but also a sufficiently remote region from the Hill sphere ($r\gg \rh$, or $r\gg1000\,\rp$), requiring the resolution of spatial scales differing by a factor of more than 1000.

Recently, \citet{dangelo03} and \citet{machida08} used the nested-grid method to resolve a wide spatial scale.
Although \citet{dangelo03} calculated a protoplanetary system in three dimensions with sufficient spatial resolution, they could not investigate the region near the protoplanet because they adopted a sink cell in the region of $r<0.1\rh$.
\citet{machida08} calculated the evolution of a protoplanetary system without a sink, and estimated the angular momentum accreted onto the system.
However, they adopted the isothermal approximation.
This approximation is considered to be valid in the circumstellar disk and areas remote from the protoplanet because these regions are optically thin due to their lower gas density.
However, this approximation is not valid in the vicinity of the protoplanet, because the gas temperature increases adiabatically as the gas density becomes high \citep{mizuno78, lunine82}.

In this study, the thermal evolution of a protoplanetary system is modelled according to  \citet{mizuno78}, and implemented into a three-dimensional nested-grid code.
Using such a code, the system is resolved from a sufficient distance from the Hill sphere ($r>10\,\rh$) to the Jovian radius ($r = 0.8\,\rp$), and the angular momentum flowing into the protoplanetary system and circumplanetary disk are investigated.
In addition, the spatial resolution in this calculation is much higher than those in \citet{dangelo03} and \citet{machida08}.
The calculations show that the angular momentum acquired by the system is independent of the thermal evolution around the protoplanet after the protoplanet grows to  $M>0.1\mj$.
A thin circumplanetary disk appears only in a compact region near the protoplanet ($r<50\rp$).
The centrifugal radius derived from the angular momentum of the protoplanetary system corresponds well to the region in which the regular satellites of present gaseous planets are distributed.
In addition, the surface density has a peak in this region.

The structure of the paper is as follows. 
The frameworks of our models are given in \S 2, and the numerical method is described in \S3. 
The numerical results are presented in \S 4.  
The discussion of circumplanetary disk and satellite formation is presented in \S5.

\section{MODEL}
\subsection{Basic Equations}

In this study, a local region around a protoplanet is considered using the shearing sheet model \citep[e.g.,][]{goldreich65}, in which the self-gravity of the disk is ignored.
In addition, no physical viscosity is included, and the numerical viscosity can be ignored because it is sufficiently small.
Thus, an inviscid gas disk model is adopted here.
The orbit of the protoplanet is assumed to be circular on the equatorial plane of the circumstellar disk. 
Local rotating Cartesian coordinates with the origin at the protoplanet are set up, in which the $x$-, $y$-, and $z$-axis are the radial, azimuthal, and vertical direction of the disk, respectively.  
The equations of hydrodynamics without self-gravity are solved:
\begin{equation}
 \dfrac{\partial \rho}{\partial t} + \nabla \cdot (\rho \, \vect{v}) = 0,\\
 \label{eq:basic-1}
\end{equation}
\begin{equation}
 \dfrac{\partial \vect{v}}{\partial t} + (\vect{v} \cdot \nabla) \vect{v} =
  - \dfrac{1}{\rho} \nabla P - \nabla \Phi_{\rm eff} - 
  2 \vect{\Omega_{\rm p}} \times \vect{v},  
  \label{eq:basic-2}
\end{equation}
 where $\rho$, $\vect{v}$, $P$, $\Phi_{\rm eff}$, and $\vect{\Omega}_{\rm p}$ are the gas density, velocity, gas pressure, effective potential, and Keplerian angular velocity of the protoplanet, respectively.  
For the gas pressure, a barotropic equation of state is adopted (for details see \S\ref{sec:eos}).
In the above equations, the curvature terms are neglected.
The Keplerian angular velocity of the protoplanet is given by
\begin{equation}
\Omega_{\rm p} = \left( \dfrac{ G\, M_{\rm c} } { \ro^3} \right)^{1/2},
\label{eq:omegap}
\end{equation}
 where $G$, $M_{\rm c}$, and $\ro$ are the gravitational constant, mass
 of the central star, and orbital radius of the protoplanet, respectively. 
The effective potential $\Phi_{\rm eff}$  is given by
\begin{equation}
\Phi_{\rm eff} = - \dfrac{\Omega_{\rm p}^2}{2}(3 x^2 - z^2) \, - \,
 \dfrac{G M_{\rm p}}{r}, 
\label{eq:phi}
\end{equation}
 where $M_{\rm p}$ and $r$ are the mass of the protoplanet, and the
 distance from the centre of the protoplanet \citep[e.g.,][]{miyoshi99}. 
The first term is composed of the gravitational potential of the central
 star and the centrifugal potential, and higher orders in $x$, $y$ and
 $z$ are neglected. 
The second term is the gravitational potential of the protoplanet.
Using the Hill radius 
\begin{equation}
 r_{\rm H} = \left( \dfrac{M_{\rm p}}{3M_{\rm c}} \right)^{1/3} \ro,
\label{eq:hill}
\end{equation} 
 equation~(\ref{eq:phi}) can be rewritten as 
\begin{equation}
\Phi_{\rm eff} = \Omega_{\rm p}^2 \left( - \dfrac{3 x^2 - z^2}{2} \, -
				   \, \dfrac{3\,r_{\rm H}^3}{r} \right). 
\label{eq:phi-eff}
\end{equation}

\subsection{Equation of State}
\label{sec:eos}
At a fixed orbital radius of the protoplanet, the circumstellar disk has an almost constant temperature \citep{hayashi85}, while the gas around the protoplanet (i.e., the gas envelope) has a higher temperature than the circumstellar disk \citep{mizuno78,mizuno80,bodenheimer86,pollack96,ikoma00}.
\citet{mizuno78} studied the structure and stability of the envelope around the protoplanet, on the assumption that the envelope is spherically symmetric, and in hydrostatic equilibrium.
They also investigated the thermal evolution of the envelope, parameterising the dust grain opacity, and determined the boundary between the isothermal and adiabatic regions.

\begin{figure}
\begin{center}
\includegraphics[width=150mm]{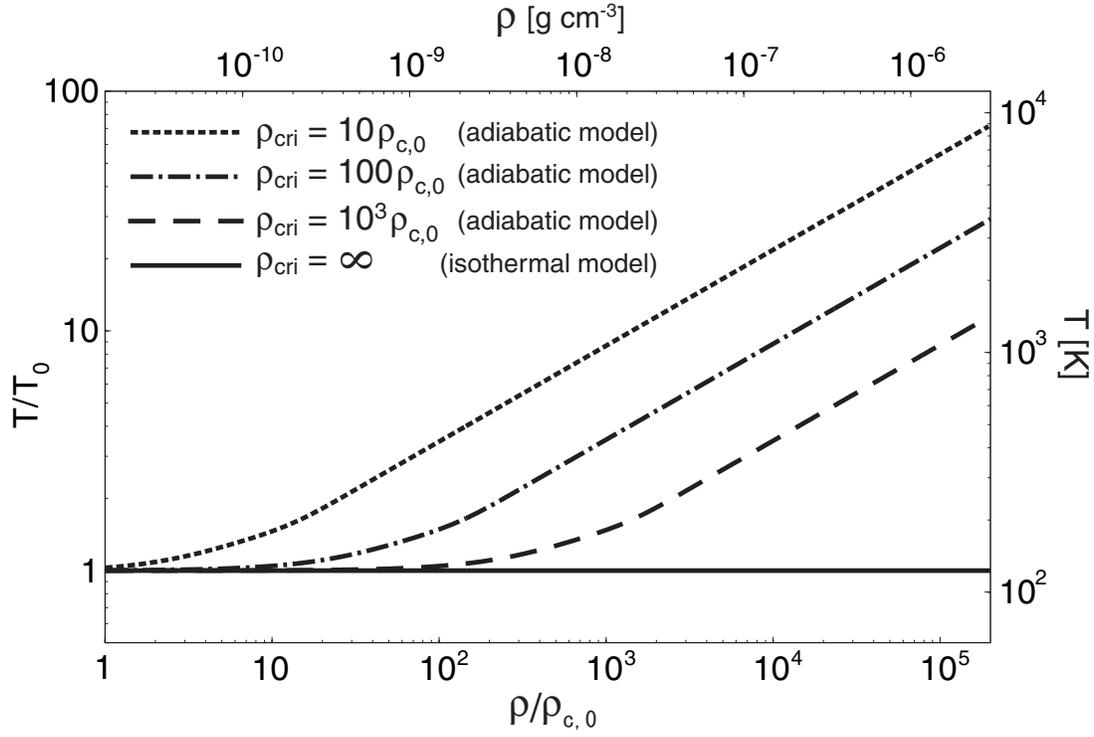}
\caption{
The density-temperature diagrams for isothermal ($\rho_{\rm cri}=\infty$) and adiabatic ($\rho_{\rm cri}$ = 10$\rho_{\rm c,0}$, 100$\rho_{\rm c,0}$, and 1000$\rho_{\rm c,0}$) models.
The right and upper axes indicate the dimensional temperature and density at the Jovian orbit, respectively.
}
\label{fig:1}
\end{center}
\end{figure}

To correctly estimate thermal evolution around the protoplanet, we need to solve the radiation hydrodynamics.
However, this has a huge computational cost.
Thus, in this study, using Figure~2 of \citet{mizuno78}, thermal evolution around the protoplanet is modelled as a function of the gas density (i.e., barotropic equation of state) : 
\begin{equation}
 P = c_{\rm s,0}^2 \rho \, \left[1-{\rm tanh}\left(\dfrac{\rho}{\rho_{\rm cri}} \right) \right]  
+ \kappa \rho^\gamma \, {\rm tanh}\left(\dfrac{\rho}{\rho_{\rm cri}} \right),
\label{eq:eos} 
\end{equation}
 where $c_{\rm s}$ is the sound speed, $\gamma$ is the adiabatic index ($\gamma=1.4$),  and 
the adiabatic constant $\kappa$ is defined as 
\begin{equation}
\kappa = c_{\rm s,0}^2 \rho_{\rm cri}^{1-\gamma},
\label{eq:kappa}
\end{equation}
where $\rho_{\rm cri}$ is  the critical density: the gas behaves isothermally in the region of $\rho < \rho_{\rm cri}$, and adiabatically in the region of $\rho > \rho_{\rm cri}$.
 In this study, $\rho_{\rm cri}$ = $\infty$ (isothermal model), 10$\rho_{\rm c,0}$, 100$\rho_{\rm c,0}$, and 1000$\rho_{\rm c,0}$ (adiabatic models) are adopted, where $\rho_{\rm c,0}$ is the initial density on the equatorial plane.   
The hyperbolic tangent (tanh) function is used to smoothly connect the first (isothermal) and second (adiabatic) terms in equation~(\ref{eq:eos}).
The thermal evolution for different $\rho_{\rm cri}$ is plotted against the gas density in Figure~\ref{fig:1}, in which the gas temperature is constant in the isothermal model ($\rho_{\rm cri}= \infty$), while it increases gradually from the initial value at $\rho\sim \rho_{\rm cri}$  in the adiabatic models ($\rho_{\rm cri}$ = 10$\rho_{\rm c,0}$, 100$\rho_{\rm c,0}$, and 1000$\rho_{\rm c,0}$).

In the standard disk model \citep{hayashi85}, the density and temperature at Jovian orbit are $\rho_{\rm c,0} =1.5\times 10^{-11}\gcm$ and  $T_0=123$\,K, respectively. 
Thus, in a model with $\rhoc = 10\,\rho_{\rm c,0}$ (the highest temperature model), the gas behaves isothermally when $\rho \ll 1.5\times10^{-10}\gcm$, while it behaves adiabatically when $\rho \gg 1.5\times 10^{-10}\gcm$.
Comparing Figure~\ref{fig:1} with Figure~2 of \citet{mizuno78}, the thermal evolution of the model with $\rho_{\rm cri}=10\, \rho_{\rm c,0}$  (Fig.~\ref{fig:1} broken line) corresponds to that for a gas envelope around a proto-Jovian planet (Fig.2 of \citealt{mizuno78}) when a gas opacity $\kappa_{\rm g}=1.0\times 10^{-2}$\,cm$^2$\,g$^{-1}$ is adopted.
\citet{mizuno78} adopted $\kappa_{\rm g}=1.0\times 10^{-4}$\,cm$^2$\,g$^{-1}$ as the most reliable parameter of a proto-Jovian planet, indicating that a more realistic gas temperature of the envelope is lower than that in the model with $\rho_{\rm cri} = 10\,\rho_{\rm c,0}$ (the dotted line of Fig.~\ref{fig:1}).
Note that the critical density $\rho_{\rm cri}$ increases as the gas opacity $\kappa_{\rm g}$ decreases.
Thus, in models with $\rhoc=10\rho_{\rm c,0}$, the thermal energy around the protoplanet may be overestimated.
On the other hand, when the isothermal equation of state is adopted, the thermal energy around the protoplanet is obviously underestimated.
Therefore, it is expected that the actual thermal evolution exists between models with $\rhoc=10\rho_{\rm c,0}$ and $\rho_c=\infty$.

\subsection{Circumstellar Disk Model}
The initial settings are similar to \citet{miyoshi99} and \citet{machida06b,machida08}.
The gas flow has a constant shear in the $x$-direction as
\begin{equation}
\vect{v_0} = ( 0,\,  -\dfrac{3}{2}\Omega_{\rm p}\, x, \,0 ).
\label{eq:shear}
\end{equation}
For hydrostatic equilibrium, the density is given by 
\begin{equation}
\rho_0 = \dfrac{\sigma_0}{\sqrt{2\pi}h} {\rm exp } \left(- \dfrac{
						    z^2}{2 h^2} \right), 
\end{equation}
 where $\sigma_0$ ($\equiv \int_{-\infty}^{\infty} \rho_0 \, dz $) is the
 surface density of the unperturbed disk. 
The scale height $h$ is related to the sound speed $c_{\rm s}$ as
 $h=c_{\rm s}/\Omega_{\rm p}$. 

In the standard solar nebular model \citep{hayashi81,hayashi85}, the
 temperature $T$, sound speed $c_{\rm s}$, and gas density $\rho_{c, 0}$ are
 given by
\begin{equation}
T = 280 \left( \dfrac{L}{L_{\odot}} \right)^{1/4}
 \left(\dfrac{\ro}{1\,{\rm AU}} \right)^{-1/2}, 
\label{eq:nebular-temp}
\end{equation}
 where $L$ and $L_{\odot}$ are the protostellar and solar luminosities,  
\begin{equation}
 c_{\rm s} = \left( \dfrac{k\,T}{\mu m_{\rm H}} \right)^{1/2} =
 1.9\times 10^4\, \left( \dfrac{T}{10\,{\rm K}} \right)^{1/2} \, \left(
		    \dfrac{2.34}{\mu} \right)^{1/2} \ \  {\rm cm\,s^{-1}}, 
\label{eq:nebular-cs}
\end{equation}
 where $\mu =2.34$ is the mean molecular weight of the gas composed
 mainly of H$_2$ and He, and   
\begin{equation}
\rho_{\rm c,0} = 1.4 \times 10^{-9} \left( \dfrac{\ro}{1\,{\rm AU}}
			    \right)^{-11/4} \ \ {\rm g}\cm. 
\label{eq:nebular-dens}
\end{equation}
When $M_c = 1\msun$ and $L=1\,L_{\odot}$ are adopted, using equations~(\ref{eq:omegap}), (\ref{eq:nebular-temp}), and (\ref{eq:nebular-cs}), the scale height $h$ can be described as
\begin{equation}
h = 5.0\times 10^{11} \left( \dfrac{\ro}{1 {\rm AU}} \right)^{5/4}  \ \ \ {\rm cm}.
\label{eq:negular-scale-height}
\end{equation}
The inverse of the angular velocity $\Omega_{\rm p}$  is described as
\begin{equation}
\Omega_{\rm p}^{-1} = 0.16 \left( \dfrac{\ro}{1 {\rm AU}} \right)^{2/3} \ \ \ {\rm yr}.
\label{eq:omegat}
\end{equation}
The non-dimensional quantities are converted into dimensional quantities using equations~(\ref{eq:nebular-temp})--(\ref{eq:omegat}).

\subsection{Scaling}
The basic equations can be normalized by unit time, $\Omega_{\rm p}^{-1}$, and unit length, $\hs$.
The density is also scalable in equations~(1) and (2) and is normalized by $\sigma_0/h$.
Hereafter, the normalized quantities are expressed with a tilde on top,
  e.g., $\tilde{x} = x/h$, $\tilde{\rho} = \rho_0/(\sigma_0/h)$, $\tilde{t}=t\,\Omega_{\rm p}$, etc. 
The non-dimensional unperturbed velocity and density are given by
\begin{equation}
 \vect{\tilde{v}} = ( 0, -\dfrac{3}{2}\tilde{x}, 0),
\end{equation}
\begin{equation}
 \tilde{\rho}_0 = 
  \dfrac{1}{\sqrt{2\pi}} {\rm exp} \left( - \dfrac{\tilde{z}^2}{2}\right).
\end{equation}
Thus, non-dimensional equations corresponding to
 equations~(\ref{eq:basic-1}), (\ref{eq:basic-2}), (\ref{eq:phi-eff}), and (\ref{eq:eos}) are 
\begin{equation}
 \dfrac{\partial \tilde{\rho}}{\partial \tilde{t}} + 
  \tilde{\nabla} \cdot (\tilde{\rho} \, \tilde{\vect{v}}) = 0,
\end{equation}
\begin{equation}
 \dfrac{\partial \tilde{\vect{v}}}{\partial \tilde{t}} + 
  (\tilde{\vect{v}} \cdot \tilde{\nabla}) \tilde{\vect{v}} =
  - \dfrac{1}{\tilde{\rho}} \tilde{\nabla} \tilde{P} - 
 \tilde{\nabla} \tilde{\Phi}_{\rm eff} - 2 \vect{\tilde{z}}  
\times \tilde{\vect{v}},  
\end{equation}
\begin{equation}
\tilde{\Phi}_{\rm eff} = 
 - \dfrac{1}{2}(3 \tilde{x}^2 - \tilde{z}^2) \, - \, \dfrac{3{\rht}^3}{\tilde{r}},
\end{equation}
\begin{equation}
\tilde {P} =  \tilde{\rho} \, \left[ 1-{\rm tanh}\left(\dfrac{\tilde{\rho}}{\tilde{\rho}_{\rm cri}}\right)  
+ \left(\dfrac{\tilde{\rho}}{\tilde{\rho}_{\rm cri}}\right)^{\gamma-1} \, {\rm tanh}\left(\dfrac{\tilde{\rho}}{\tilde{\rho}_{\rm cri}} \right) \right],
\end{equation}
 where $\tilde{\vect{z}}$ is a unit vector directed to the $z$-axis.
The gas flow is characterized by two parameters, the non-dimensional Hill radius $\rht=\rh/h$, and critical density $\tilde{\rho}_{\rm cri}$.
In this study, the Hill radii $\rht =0.29-1.36 $  are adopted.
As a function of the orbital radius and the mass of the central star, the parameter $\rht$ is related to the actual mass of the protoplanet in units of Jovian mass $M_{\rm J}$ as 
\begin{equation}
 \dfrac{M_{\rm p}}{M_{\rm J}} = 
 0.12 \left( \dfrac{M_{\rm c}}{1\msun} \right)^{-1/2} 
 \left( \dfrac{\ro}{1\,{\rm AU}} \right)^{3/4} \, \rht^3.
\label{eq:mass-to-hill}
\end{equation}
For example, in the model with $\rht = 1.0$,  $\ro=5.2$\,AU and
 $M_{\rm c} = 1\,\msun$, the protoplanet mass is 
 $M_{\rm p} = 0.4 M_{\rm J}$ (model M04 in Table~\ref{table:table1}). 
In the parameter range of $\rht =0.29-1.36 $,  at Jovian orbit ($\ro=5.2$\,AU), protoplanets have masses of $0.01-1\mj$.
Model names, non-dimensional Hill radii $\rht$, masses of protoplanets at Jovian (5.2\,AU) and Saturnian orbits (9.6\,AU) and critical densities are listed in Table~1.
Model names consist of two parts: the mass of the protoplanet at Jovian orbit and critical density.
For example, model M001A3 has parameters of ($M_{\rm p}$, $\rho_{\rm cri}$) = (0.01$\mj$, $10^3\,\rho_{\rm c,0}$).

\section{NUMERICAL METHOD}
\subsection{Nested-Grid Method}
\label{sec:nested-grid}
To investigate the formation of a circumplanetary disk in a circumstellar or protoplanetary disk, it is necessary to cover a large dynamic range of spatial scale.
Using the nested-grid method \citep[for details, see][]{machida05, machida06a}, the regions near (using grids with higher spatial resolution) and remote from (using grids with lower spatial resolution) the protoplanet are covered. 
Each level of a rectangular grid has the same number of cells ($ = 64 \times 128 \times 16 $), but cell width $\Delta \tilde{s}(l)$ depends on the grid level $l$. 
The cell width is reduced by 1/2 with increasing grid level ($l \rightarrow l+1$).
We use 8 grid levels ($l_{\rm max}=8$).
The box size of the  coarsest grid, $l=1$, is $(\tl{L}_x, \tl{L}_y, \tl{L}_z) = (12, 24, 3)$, and that of the finest grid, $l=8$, is $ (\tl{L}_x, \tl{L}_y, \tl{L}_z) = (0.09375, 0.1875, 0.0234)$. 
The cell width in the coarsest grid, $l=1$, is $\Delta \tilde{s} = 0.1875$, and it decreases with $\Delta \tilde{s}=0.1875/2^{l-1}$ as the grid level $l$ increases.
Thus, the finest grid has  $\Delta \tilde{s}(8)\simeq 1.46\times10^{-3}$.
The fixed boundary condition in the $\tilde{x}$- and $\tilde{z}$-direction, and the periodic boundary condition in the $\tilde{y}$-direction are adopted.

The local simulation is not appropriate to treat the gap formation because of the radial boundary \citep{miyoshi99,tanigawa02}. 
The gap property (the gap depth and width) depends on the size of the simulation box. 
Although the density gap may affect the mass accretion rate onto the protoplanet, it does not affect the formation of the circumplanetary disk, because the formation of circumplanetary disk depend only on the size of Hill sphere \cite[see,][]{machida08}.

\subsection{Sink Cell and Smoothing Length}
In the finest grid ($l_{\rm max}=8$), the cell width is $\Delta \tl{s} = 1.46\times 10^{-3}$.
In real units, when the protoplanet is located at 5.2\,AU, the cell width corresponds to $\Delta s = 5.7\times 10^{9}$\,cm, or 0.8 times the Jovian radius.
In each model, the evolution of the protoplanetary system was calculated adopting the sink.
The radius of the sink is $\tilde{r}_{\rm sink}=3.53\times 10^{-3}$ or twice the Jovian radius at Jovian orbit. 
During the calculation, the gas from the region inside the sink radius is removed in each time step.
\citet{machida08} investigated the effect of the sink by parameterising the sink radius, and found that the circumplanetary disk and angular momentum of the protoplanetary system can be appropriately estimated when the sink radius is sufficiently smaller than the Hill radius ($r\ll0.1\,\rh$).

The smoothing length for the gravitational potential of the protoplanet is not explicitly adopted.
In numerical settings, the physical quantities are defined at the cell centre, while the origin (protoplanet's position) is defined at the cell boundary.
Thus, since the region inside $ \tilde{r} < \tilde{r}_s \equiv  \sqrt{3} \Delta \tilde{s}(l_{\rm max})/2 $ has a uniform gravitational potential.
At Jovian orbit, $r_s$ has a $0.7$ times the Jovian radius ($\tilde{r}_s = 1.26 \times 10^{-3}$ or $r_s = 4.9\times10^9$\,cm). 

\section{Results}
\label{sec:results}
\subsection{Models with Jovian-Mass Protoplanet}
In this subsection, to investigate the relationship between thermal effects and the structure and angular momentum of the protoplanetary system in detail, only models with a Jovian-mass protoplanet at Jovian orbit (models M1A1, M1A2, M1A3, M1I) are shown. 
\label{sec:jovian-mass}

\subsubsection{Large Scale Structure and Density Distribution}
\label{sec:typical}
In each model, the evolution of the protoplanetary system was calculated for $\deft \ge 600$ (or $\ge$100 orbits), in which the flow around the protoplanet reached a steady state in a short timescale of $\deft \sim 6-10$ ($\sim 1-10$ orbit).
\footnote{
The non-dimensional time unit $\deft$ can be converted into the orbital period as $\deft/(2\pi)$. 
}
Previous studies also show a steady state in a short timescale \citep[e.g.,][]{miyoshi99,tanigawa02,machida06b}.
In this study, for safety, the evolution of the protoplanetary system was calculated for a sufficiently long time ($\tilde{t}\ge 600$) in all models.
Since the evolution of the system until the steady state was already shown in \citet{machida08} in detail, only the evolution after the steady state is shown below.

The left panel of Figure~\ref{fig:2} shows the density (colour) and velocity distribution (arrows) in the equatorial plane ($\tilde{z}=0$) at $\deft=637.23$ for model M1I, in which the gas behaves isothermally.
Model M1I has a mass of $1\mj$ at the Jovian orbit $a_{\rm p} = 5.2$\,AU.
To stress the structure inside the Hill radius, represented by the dashed line, only the grids of $l=3-6$ are plotted in this panel.
Note that the grid size of $l=3$ is $(\tl{L}_x, \tl{L}_y, \tl{L}_z)$ = (3, 6, 0.75), while the boundary of the outermost grid ($l=1$) is located at  ($x$, $y$, $z$)= $(12, 24, 3)$.
This panel shows the shocks (crowded contours near the Hill radius) in the upper right and lower left region against the protoplanet (or the centre).
Inside the Hill radius, a non-axisymmetric pattern appears.
These features  are also seen in previous studies \citep{kley99, miyoshi99, lubow99, kley01, dangelo02, dangelo03}.

The right panel of Figure~\ref{fig:2} shows the density (colour) and velocity distribution (arrows) in the equatorial plane ($\tilde{z}=0$) at $\tilde{t}=637.73$ for model M1A1, in which the temperature increases gradually as the gas density increases, as shown in Figure~\ref{fig:1}.
Figure~\ref{fig:2} shows that the structures in models M1I and M1A1 are nearly identical, while the shock in model M1A1 appears in a more remote region from the protoplanet than in model M1I.
In addition, model M1A1 has a more axisymmetric structure near the protoplanet ($\tilde{r}<0.5$) than model M1I.
These differences are caused by thermal effects around the protoplanet.
The larger thermal pressure forms a more spherical structure because the thermal pressure gradient force is isotropic.

Figure~\ref{fig:3} shows the density (colour) and velocity distribution (arrows) in the $\tilde{y}=0$ plane for models M1I, M1A3, M1A2 and M1A1, in which three levels of the grid ($l=6$, 7 and 8) are superimposed.
This figure indicates that models with higher gas temperature form relatively thick envelopes.
The envelopes of models M1I and M1A3 have a height-to-radius ratio of $H/R \sim1/10$ in the region of $\tilde{\rho}>10^3$, while that of model M1A1 has $H/R \sim 1/5$ in the region of $\tilde{\rho}>10^3$ (see, the contours for $\tilde{\rho}=10^3$ in  Fig.~\ref{fig:3}).
However, even in model M1A1, the envelope in the region of $\tilde{\rho}>10^4$ is sufficiently thin ($H/R\sim1/10$). 
Thus, a thin disk forms close to the protoplanet in each model, while the protoplanetary systems have different envelope thicknesses in the region far from the protoplanet for models with different equations of state.
In adiabatic models, since the gas temperature increases with the gas density as shown in Figure~\ref{fig:1}, the thermal effect is more significant close to the protoplanet.
Despite this, the structural difference is clearer in the region far from the protoplanet (i.e., the low-density region).
This indicates that the protoplanet's gravity and centrifugal force significantly dominate the thermal pressure gradient force close to the protoplanet.

\subsubsection{Angular momentum of Protoplanetary System}
\label{sec:typical2}
Figure~\ref{fig:4} shows the distribution of the average specific angular momentum  $\tilde{j}_r$ against the distance from the protoplanet $\tilde{r}$ for models M1I, M1A1, M1A2 and M1A3.
The average specific angular momentum $\tilde{j}_r$ is defined as
\begin{equation}
\tl{j}_r = \dfrac{\tl{J}_r}{\tl{M}_r}, 
\label{eq:jr}
\end{equation}
where the mass 
\begin{equation}
\tl{M}_r = \int^{r}_0  \, 4\pi \tl{r}^2 \tl{\rho}\, d\tl{r},
\label{eq:mr}
\end{equation}
and angular momentum 
\begin{equation}
\tl{J}_r  = \int^{r}_0  4\pi \tl{r}^2 \tl{\rho} \, \tl{\varpi} \tl{v}_{\phi}\,  d\tl{r},
\label{eq:Jr}
\end{equation}
are integrated from the centre ($\tl{r}=0$) to a distance $r$.
The solid line in Figure~\ref{fig:4} shows the distribution of $\tilde{j}_r$ for the isothermal model (M1I1).
This distribution corresponds well to that of \citet{machida08}, in which they investigated the accretion of angular momentum into a protoplanetary system using an isothermal equation of state without a sink cell,  and found from the Jacobi energy, Kepler velocity and distribution of mass and angular momentum around the protoplanet that the angular momenta bound by protoplanetary system are limited in the region of $\tilde{r} \le 0.5\rht-1\rht$.
As shown in Figure~\ref{fig:4} and \citet{machida08}, in each model, the average specific angular momentum increases from the centre to a peak around the Hill radius, and drops sharply just outside the Hill radius.
The drop indicates that the angular momentum becomes negative at $\tilde{r}>\rht$.
Thus, the rotational direction is opposite in region inside and outside the Hill radius.
As shown in \citet{sekiya87}, \citet{miyoshi99}, and \citet{tanigawa02}, a protoplanet formed by gas accretion in the circumstellar disk has a prograde spin, and thus it has a positive (specific) angular momentum.
On the other hand, gas far outside the Hill sphere seems to rotate retrogradely against the protoplanet because it rotates with nearly Keplerian velocity with respect to the central star [$v=-(3/2)\, \Omega_{\rm p}\,x$; see eq.~(\ref{eq:shear})].
As a result, gas inside the Hill radius has a positive angular momentum, while that outside has a negative angular momentum against the protoplanet.

Figure~\ref{fig:4} shows little difference in  average specific angular momentum between the isothermal (M1I) and adiabatic (M1A1, M1A2, and M1A3) models in the region of $\tilde{r}<0.1$, while there are no differences in the region of $\tilde{r}>0.1$.
Thus, the gas envelope near the protoplanet is slightly influenced  by thermal effects, while differences in thermal energy do not affect the evolution of angular momentum in the region of $\tilde{r}>0.1$.
Since the angular momentum bound by the protoplanet is mainly distributed in the region of $0.5\,\rht\le \tilde{r} \le 1\rht$ \citep{machida08}, the different equations of state (or different gas temperatures) hardly affect the total angular momentum acquired by a protoplanetary system from the circumstellar disk.
\citet{machida08} noted that gravitational energy dominates thermal energy inside the Hill radius when the mass of the protoplanet exceeds $M_{\rm p}>0.08\mj$ at Jovian orbit.
Since models M1I, M1A1, M1A2 and M1A3 have a mass of $1\mj$ at Jovian orbit,  the gravitational energy dominates the thermal energy greatly.
Thus, it is natural that different thermal energies hardly affect the evolution of the protoplanetary system and the acquisition process of angular momentum in these models.

\subsection{Relation Between Angular Momentum and the Protoplanet's Mass}
\label{sec:am-deos}
In \S\ref{sec:jovian-mass}, we examined only the evolution of models having protoplanets with $1\mj$.
In this subsection, the evolution of angular momentum for models with different protoplanetary masses is described.

The average specific angular momenta derived from all models are plotted against the protoplanet's mass in Figure~\ref{fig:5}.
The average specific angular momenta are estimated in the region of $\tilde{r} <  \rht/2$.
Note that \cite{machida08} showed that the angular momentum bound by the protoplanetary system is distributed in the range of $1/2\,\rht\le \tilde{r} \le \rht$ and has almost the same value in this range \citep[for details, see Fig.~11 of][]{machida08}.
Figure~\ref{fig:5} indicates that the system has  almost the same average specific angular momentum in models with the same protoplanetary mass but different equations of state when the mass of the protoplanet exceeds $0.1\mj$.
Thus, although the protoplanetary systems have different thermal energies around the protoplanet, they acquire almost the same specific angular momentum from the circumstellar  disk when $M_{\rm p}> 0.1\mj$.
On the other hand, when the protoplanetary mass is smaller than $0.1\mj$, the specific angular momenta for adiabatic models are slightly smaller than that for the isothermal model.
Thus, the thermal pressure impedes the acquisition of angular momentum from the circumstellar disk when $M<0.1\mj$.
However, since a large fraction of gas and angular momentum falls into the protoplanetary system after the mass of the protoplanet becomes comparable to the present masses of gas giant planets, a difference in specific angular momentum in the lower-mass phase ($M<0.1\mj$) hardly affects the subsequent evolution.

The thin solid line in Figure~\ref{fig:5} denotes $j \propto M$.
\citet{machida08} showed that, using the isothermal equation of state, the specific angular momentum is fitted by $j \propto M$ when the protoplanet has a mass of $0.1\mj< M<1\mj$.
Figure~\ref{fig:5} shows that the average specific angular momentum can be fitted by $j \propto M$ even when the adiabatic equation of state is adopted.
As shown in \citet{machida08}, in real units, the average specific angular momentum of a protoplanetary system is approximated as 
\begin{eqnarray}
j_{\rm lm} &=& 7.8 \times 10^{15} \left(\dfrac{M_{\rm p}}{M_{\rm J}} \right)
\left(\dfrac{\ro}{1\,{\rm AU}}  \right)^{7/4} \jcm, 
\label{eq:fit} 
\end{eqnarray}
when the protoplanet has a mass of $M_{\rm p}<1\mj$.
Although \citet{machida08} estimated the angular momentum in the isothermal regime, equation~(\ref{eq:fit}) is valid even in the adiabatic regime as shown in Figure~\ref{fig:5}.

\citet{lissauer95} analytically estimated the specific angular momentum $j_{\rm ana}$ of a proto-planetary system (see, also \citealt{stevenson86} and \citealt{mosqueira03}) as
\begin{equation}
j_{\rm ana} = \dfrac{1}{4} \Omega_{\rm p} \rh^2.
\label{eq:j-ana}
\end{equation}
Using equations~(\ref{eq:omegap}) and (\ref{eq:hill}), equation~(\ref{eq:j-ana}) can be rewritten as
\begin{equation}
j_{\rm ana} = 5.2 \times 10^{16} \left( \dfrac{M_{\rm p}}{1 \mj} \right)^{2/3} 
\left( \dfrac{\ro}{1\, {\rm AU}} \right)^{1/2} \jcm.
\label{eq:j-ana2}
\end{equation}
The different power of $M_{\rm p}$ and $\ro$ between equations~(\ref{eq:fit}) and (\ref{eq:j-ana2}) is due to the disk model adopted.
The standard disk model [see eqs.~(\ref{eq:nebular-temp})--(\ref{eq:negular-scale-height})] is adopted in this study, while equation~(\ref{eq:j-ana2}) is derived assuming only a Keplerian rotating disk in which the thermal pressure of the circumstellar disk is ignored.
Despite this difference, equation~(\ref{eq:j-ana2}) is quantitatively the same as equation~(\ref{eq:fit}).
For example, with a proto-Jovian planet ($M_{\rm p} = 1\mj$) at Jovian orbit ($\ro$=5.2\,AU), the protoplanetary system has a specific angular momentum of $j_{\rm lm, J} = 1.4\times 10^{17}\jcm$ in equations~(\ref{eq:fit}), and $j_{\rm ana, J} = 1.2\times 10^{17}\jcm$ in equation~(\ref{eq:j-ana}).
With a proto-Saturnian planet ($M_{\rm p, J} = 0.3\mj$) at Saturnian orbit ($\ro$=9.6\,AU), the specific angular momenta are $j_{\rm lm, S} = 1.2\times 10^{17}\jcm$ in equation~(\ref{eq:fit}), and  $j_{\rm ana, J} = 7.2\times 10^{16}\jcm$ in equation~(\ref{eq:j-ana2}).
Equations~(\ref{eq:fit})-(\ref{eq:j-ana2}) indicate that the specific angular momentum of a protoplanetary system is determined only by the protoplanet's mass (or the size of the Hill sphere) at a fixed orbit.
Since the gravitational energy dominates the thermal energy inside the Hill sphere when $M>0.08\mj$ at Jovian orbit as described in \citet{machida08}, the thermal effect can be ignored in the angular momentum acquisition process when the protoplanet's mass is  $M>0.08\mj$.
Thus, it is natural that the angular momentum flowing onto the system is controlled only by the protoplanet's mass.

The resulting angular momentum of the protoplanetary system can be estimated by integrating equation~(\ref{eq:fit}) using mass up to the present values for gas giant planets.
As a result, the Jovian and Saturnian protoplanetary systems have $\sim30$ and $\sim50$ times larger angular momentum than the present gas giant planet systems  \citep[the central gas giant planet + satellites, for details see][]{machida08}.


\subsection{Circumplanetary Disk around Proto-Jupiter and Saturn}
Figure~\ref{fig:6} shows the density distribution on the $y=0$ plane (upper) and surface density along the $z$-axis (lower) around proto-Jupiter (left; model M1A2) and proto-Saturn (right; model M02A2).
Model M1A2 (left panels) has a parameter of $\rht=1.36$, which corresponds to a Jovian mass (1$\mj$) at Jovian orbit (5.2\,AU), while model M02A2 (right panels) has $\rht=0.8$, which corresponds to a Saturnian mass ($1M_{\rm s}$) at Saturnian orbit (9.6\,AU).
In addition, these models have  $\rhoc=100\rho_{\rm c,0}$, which is the most realistic parameter when a Jovian mass is adopted \citep[see][]{mizuno78}.
Since the evolution of protoplanetary systems was calculated in non-dimensional units,  the calculation results can be scaled in real units at an arbitrary orbit.
Figure~\ref{fig:6} adopts the units at Jovian (left) and Saturnian (right) orbit.  
The dotted lines in Figure~\ref{fig:6} represent 10, 30 and 50 times the planet's radius; a Jovian radius is adopted for model M1A2 (left panel), while a Saturnian radius is adopted for model M02A2 (right panel).

The contours in the upper panels of Figure~\ref{fig:6} show a thin disk close to the protoplanet ($r \le 50\,r_{\rm p}$), while the gas envelope far from the protoplanet ($r>50\,r_{\rm p}$) has a thick torus-like density distribution.
In the region far from the protoplanet ($r>50\,\rp$), the disk is flared, because the circumplanetary disk connects smoothly to the circumstellar disk, which is thicker than the circumplanetary disk.
Owing to rapid rotation near the protoplanet, the disk thins as it approaches the protoplanet.
Figure~\ref{fig:7} also shows the density distributions around the protoplanets in a bird's-eye view, but at a larger scale than Figure~\ref{fig:6}.
Each surface in Figure~\ref{fig:7} represents an iso-density surface of $\rho=10^3\,\rho_{\rm c,0}$ (red), $100\,\rho_{\rm c,0}$ (orange) and $10\,\rho_{\rm c,0}$ (green and blue). 
The density distribution in the $x=0$, $y=0$ and $z=0$ plane is projected onto each wall surface.
Figure~\ref{fig:7} shows that the central region sags in the centre of a concave structure, and a thin disk is formed around the protoplanet.
In addition, a butterfly-like structure is also seen on each wall surface.
These structures are considered to be formed by the rapid rotation of the central circumplanetary disk.
The thin disk near the protoplanet (red  surface) is enclosed by a torus-like thick disk (orange and blue surfaces).
Figures~\ref{fig:6} and \ref{fig:7} indicate that a thin disk appears only in a compact region of $r<50\,\rp$, and disk thickness becomes drastically large in the region of $r>50\,r_{\rm p}$.
These features are considered to result from the angular momentum of the protoplanetary system.

Using the centrifugal radius, the radial extent of the circumplanetary (thin) disk can be related to the angular momentum of the protoplanetary system. 
Under the assumption that the centrifugal force is balanced by the gravity of the protoplanet, the centrifugal radius $r_{\rm cf}$ is denoted  as 
\begin{equation}
r_{\rm cf} = \dfrac{j_{\rm r}^2}{GM_{\rm p}}.
\label{eq:cent}
\end{equation}
When the specific angular momentum $j_{\rm r}$ and mass of the protoplanet $M_{\rm p}$ are given in equation~(\ref{eq:cent}), the centrifugal radius can be estimated.
Using the specific angular momentum of the protoplanetary system given by equation~(\ref{eq:fit}), the centrifugal radius for the proto-Jovian disk with $M_{\rm p} = 1 \mj$ and $\ro =5.2$\,AU is
\begin{equation}
r_{\rm cf, J}  = 1.5 \times10^{11} \ \ {\rm cm},
\label{eq:rcent-j}
\end{equation}
which is 21 times the Jovian radius ($r=21\,r_{\rm Jup}$, where $r_{\rm Jup} = 7.1\times 10^9\cm$) or 0.028 times the Jovian Hill radius ($r=0.028\,r_{\rm H,Jup}$, where $r_{\rm H,Jup}=5.3\times10^{12}$cm).
In the same way, the centrifugal radius of the proto-Saturnian disk is estimated as
\begin{equation}
r_{\rm cf, S}  = 4.0\times10^{11} \ \ {\rm cm},
\label{eq:rcent-s}
\end{equation}
which is 66 times the Saturnian radius ($r=66\,r_{\rm Sat}$, $r_{\rm Sat}=6.0\times10^9$\,cm) or 0.061 times the Saturnian Hill radius ($r=0.061\,r_{\rm H,Sat}$, where $r_{\rm H,Sat}=6.5\times10^{12}$cm).
These values are the same as the results of \citet{machida08}, in which the evolution of a protoplanetary system was investigated under the isothermal approximation with coarser spatial resolution than this study.
Observations showed that the regular satellites around Jupiter and Saturn are distributed in a compact region in close proximity to the planet ($r\le 50\,\rp$), and are on prograde orbits near the equatorial plane.
The Galilean satellites, which are regular  satellites of Jupiter, are distributed in the range of $6\,r_{\rm Jup}\le r \le 26\,r_{\rm Jup}$, which is consistent with the centrifugal radius of equation~(\ref{eq:rcent-j}).
The Saturnian representative regular satellites (Mimas, Enceladus, Tethys, Dione, Rhea, Titan,  and Iapetus) are distributed in the range of $3\, r_{\rm Sat} \le  r \le 60\, r_{\rm Sat}$, which corresponds well to equation~(\ref{eq:rcent-s}).
In addition, the lower panels of Figure~\ref{fig:6} show that the surface density has a peak in the region of $10\,r_{\rm p}<r<50\,r_{\rm p}$.
This is as a result of gas accumulation in this annulus because of the centrifugal barrier.
Thus, regular satellites may form in this compact region.

\section{Circumplanetary Disk and Regular Satellites}
After the formation of the circumplanetary gas disk, satellites may form in the disk.
Although there are many scenarios for satellite formation \citep[e.g.,][]{stevenson86}, regular satellites around gas giant planets are believed to form in the gaseous disk as planets form in the circumstellar  disk \citep[e.g.,][]{stevenson86, korycansky91, canup02}.
Regular satellites are expected to form in the circumplanetary disk simultaneously with planets: the dust component condenses into a dust layer that fragments into satellitesimals in a thin disk near the protoplanet, and these satellitesimals merge to form satellites. 
\citet{lunine82} suggested the minimum mass subnebular (MMSN) model, in which the mass of the present regular Jovian satellite system is reconstructed assuming the dust-gas-ratio of the solar nebula. 
However, in this model, there are some difficulties for regular satellite formation.
For example, planetesimals $\sim1$\,km in size fall into a central protoplanet within $\sim1000$\,yr due to the gas drag \citep{stevenson86}.
In addition, the disk itself is expected to be dissipated by viscous spreading within $\sim100$\,yr \citep{stevenson86}.
To overcome these difficulties, \citet{canup02} suggested the gas-starved disk model, in which the circumplanetary disk has several orders of magnitude lower surface density than in the MMSN disk model.
In this scenario, solids delivered to the disk build up over many disk viscous cycles, resulting in a greatly reduced gas-to-solids ratio during the final stages of satellite accretion.

At the Callisto orbit ($r\simeq26\,r_{\rm Jup}$), the MMSN disk model has $\sigma\simeq 7.1\times10^5\scm$ \citep{lunine82}, while the gas-starved disk model has $\sigma\simeq 100\scm$ \citep{canup02}.
Note that the gas-starved model assumes a very late stage of gas accretion onto Jupiter, in which the gas flows to the protoplanet very slowly.
In this study, as shown in Figure~\ref{fig:6}, the surface density at the Cellist orbit is $\sigma \simeq 10^4\scm$ which is an intermediate value between those of the MMSN and gas-starved disk models.
The Jovian representative regular satellites (Galilean satellites) are distributed in the region of $5.9\,r_{\rm p}<r<26.3\,r_{\rm p}$.
This region is sufficiently resolved in this calculation, since the cell width and sink radius are $0.8\, \rp$ and $2 \, \rp$, respectively.
However, the inner boundary condition (i.e., the sink) may be inappropriate for investigating a circumplanetary disk close to the protoplanet.

The sink cell is adopted to calculate the evolution of the protoplanetary system for a long duration.
Without the sink, such calculation is impossible, because gas falls into the central region with a velocity much higher than the sound speed, and the time step becomes quite small.
To properly estimate the circumplanetary disk in very close proximity to the protoplanet ($r<5\,\rp$), the protoplanet model (density and temperature) must be set at the centre.
However, such treatment is beyond our scope: we cannot know the equation of state of the interior region of the protoplanet.
Note that it is impossible to perform a calculation for the protoplanet itself, because it requires much greater spatial resolution.

However, it is expected that the properties of the circumplanetary disk do not change qualitatively irrespective of the treatment of the inner boundary.
\citet{machida08} showed that the specific angular momentum and configuration of the disk are determined by the size of the Hill radius (or gravitational potential of the protoplanet).
This study showed that the specific angular momentum of a protoplanetary system is independent of the thermal evolution around the protoplanet after the protoplanet's mass exceeds $0.1 \mj$, indicating that the properties of the circumplanetary disk are determined only by the outer boundary (i.e. the size of the Hill radius).
On the other hand, the region in very close proximity to the protoplanet is slightly affected by thermal evolution, as shown in Figure~\ref{fig:4}.
Since regular satellites are believed to form  in this region, more advanced calculations are necessary to determine the circumstellar disk model closely related to satellite formation.

\section*{Acknowledgments}
We thank ~T. Matsumoto for contributions to the nested-grid code.
We have greatly benefited from discussion with ~S. Inutsuka, ~K. Kokubo, and  ~S. Ida.
Numerical computations were carried out on VPP5000 at Center for Computational Astrophysics, CfCA, of National Astronomical Observatory of Japan.
This work is supported by the Grants-in-Aid from MEXT (1607702 and 18740104).

\clearpage
\begin{table}   
\caption{Model parameters}
\label{table:table1}
\begin{center}
\begin{tabular}{cccccccccccccc}
\hline
Model & $\rht$   &  {\scriptsize $M_{\rm p}$}$^*$  {\scriptsize (5.2 AU)} &  
{\scriptsize $M_{\rm p}$}$^*$  {\scriptsize (9.6 AU)}
&  $\tilde{\rho}_{\rm cri}$

\\
\hline
M001 (M001A1, M001A2, M001A3, M001I) & 0.29   & 0.01 & 0.016 & (10, 100, 1000, $\infty$) \\
M005 (M005A1, M005A2, M005A3, M005I) & 0.5    & 0.05 & 0.081 & (10, 100, 1000, $\infty$) \\
M01 (M01A1, M01A2, M01A3, M01I)      & 0.63   & 0.1  & 0.16  & (10, 100, 1000, $\infty$) \\
M02 (M02A1, M02A2, M02A3, M02I)      & 0.8    & 0.2  & 0.35  & (10, 100, 1000, $\infty$) \\
M04 (M04A1, M04A2, M04A3, M04I)      & 1.0    & 0.4  & 0.65  & (10, 100, 1000, $\infty$) \\
M06 (M06A1, M06A2, M06A3, M06I)      & 1.15   & 0.6  & 1.0   & (10, 100, 1000, $\infty$) \\
M08 (M08A1, M08A2, M08A3, M08I)      & 1.26   & 0.8  & 1.3   & (10, 100, 1000, $\infty$) \\
M1 (M1A1, M1A2, M1A3, M1I)          & 1.36   & 1    & 1.7    & (10, 100, 1000, $\infty$) \\
\hline
\end{tabular}
\end{center}
* {in unit of Jupiter mass  $M_{\rm J}$}
\end{table}

\clearpage

\clearpage
\begin{figure}
\begin{center}
\includegraphics[width=160mm]{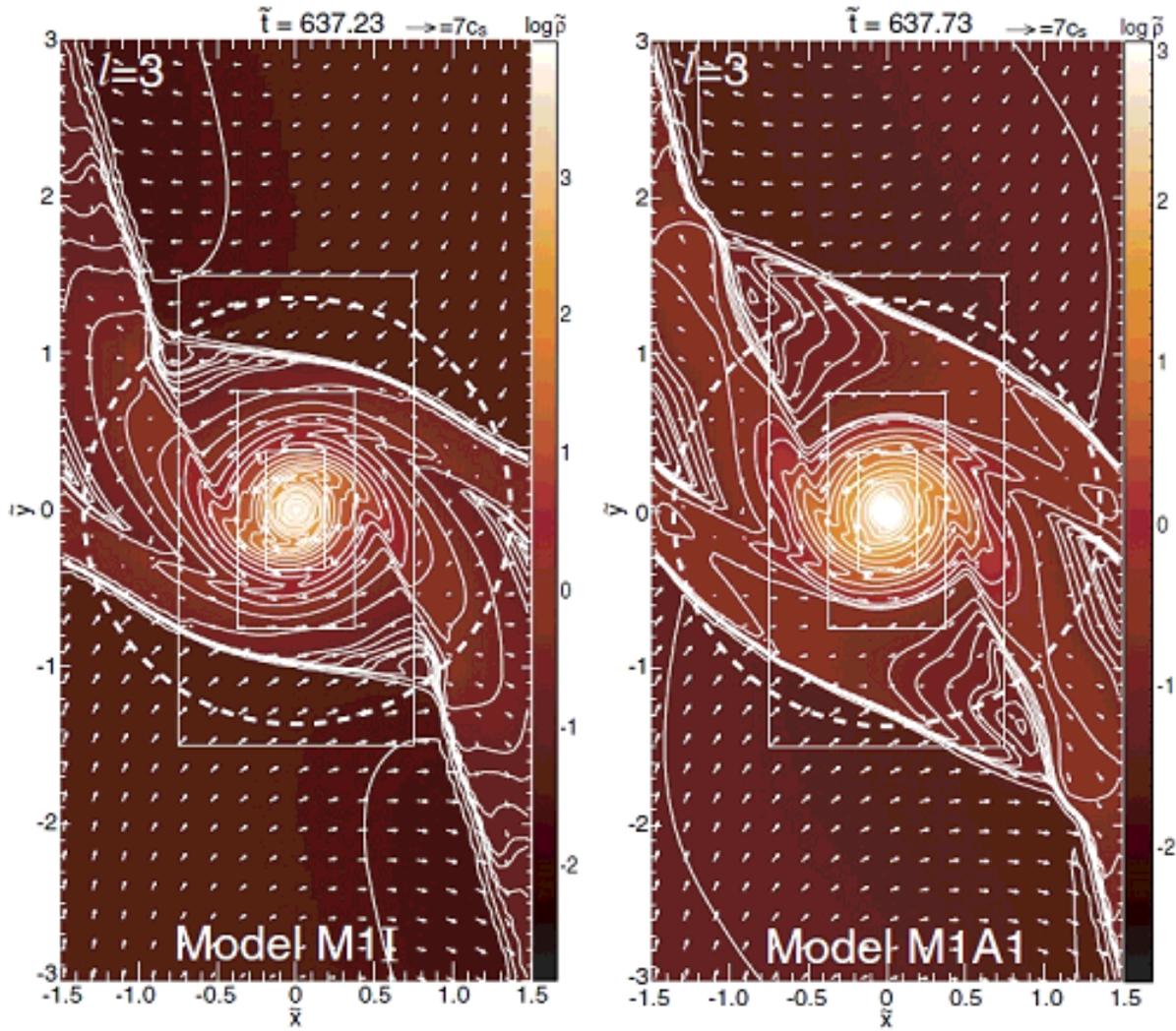}
\caption{
Density ({\it colour scale}) and velocity distribution ({\it arrows}) on the cross section in the $\tl{z}=0$ plane for models M1I (left) and M1A1 (right).
Level of the outermost grid is denoted in the upper left corner.
Elapsed time $\deft$ and velocity scale are given at the top of each panel.
Dashed circle indicates the Hill radius.
}
\label{fig:2}
\end{center}
\end{figure}

\clearpage
\begin{figure}
\begin{center}
\includegraphics[width=160mm]{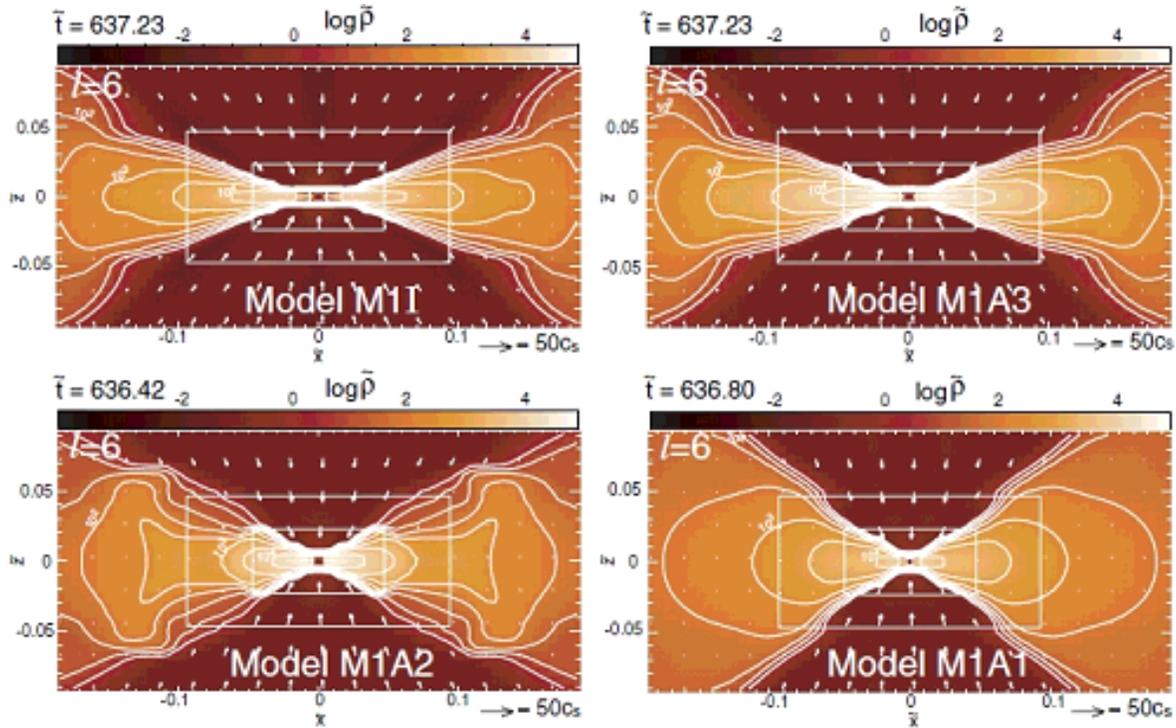}
\caption{
Density ({\it colour scale}) and velocity distribution ({\it arrows}) on the cross section in the $\tl{y}=0$ plane for models M1I (top left), M1A3 (top right), M1A2 (bottom left), and M1A1 (bottom right).
Level of the outermost grid is denoted in the upper left corner.
Elapsed time $\deft$ is given at the top of each panel.
The velocity scale in units of the sound speed is denoted in the bottom of each panel.
}
\label{fig:3}
\end{center}
\end{figure}

\clearpage
\begin{figure}
\begin{center}
\includegraphics[width=160mm]{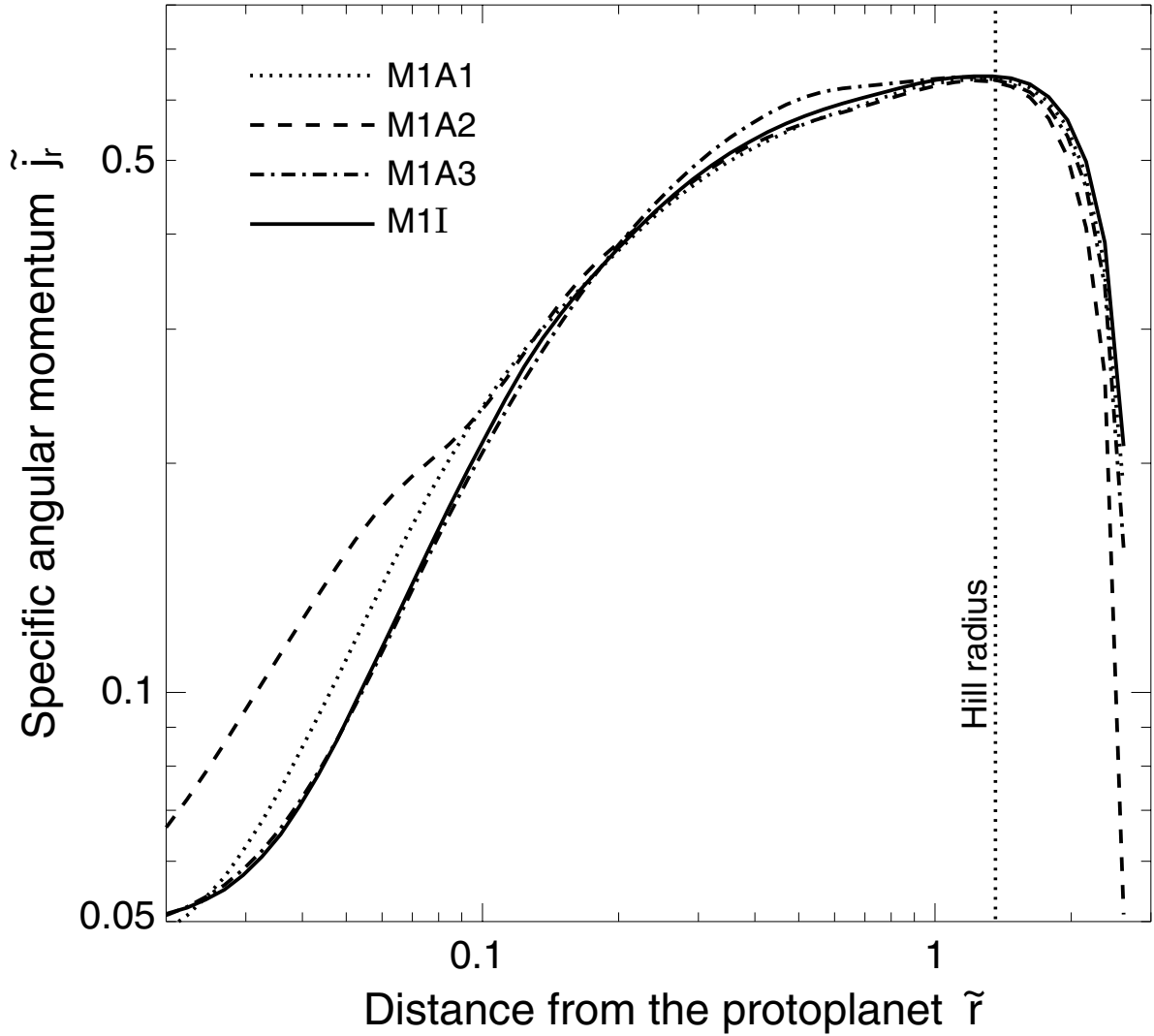}
\caption{
Average specific angular momentum $\tilde{j}$ against the distance $\tilde{r}$ from the protoplanet for models M1I, M1A1, M1A2 and M1A3.
The vertical dotted line represents the Hill radius, $\tilde{r} = 1.36$.
}
\label{fig:4}
\end{center}
\end{figure}

\clearpage
\begin{figure}
\begin{center}
\includegraphics[width=160mm]{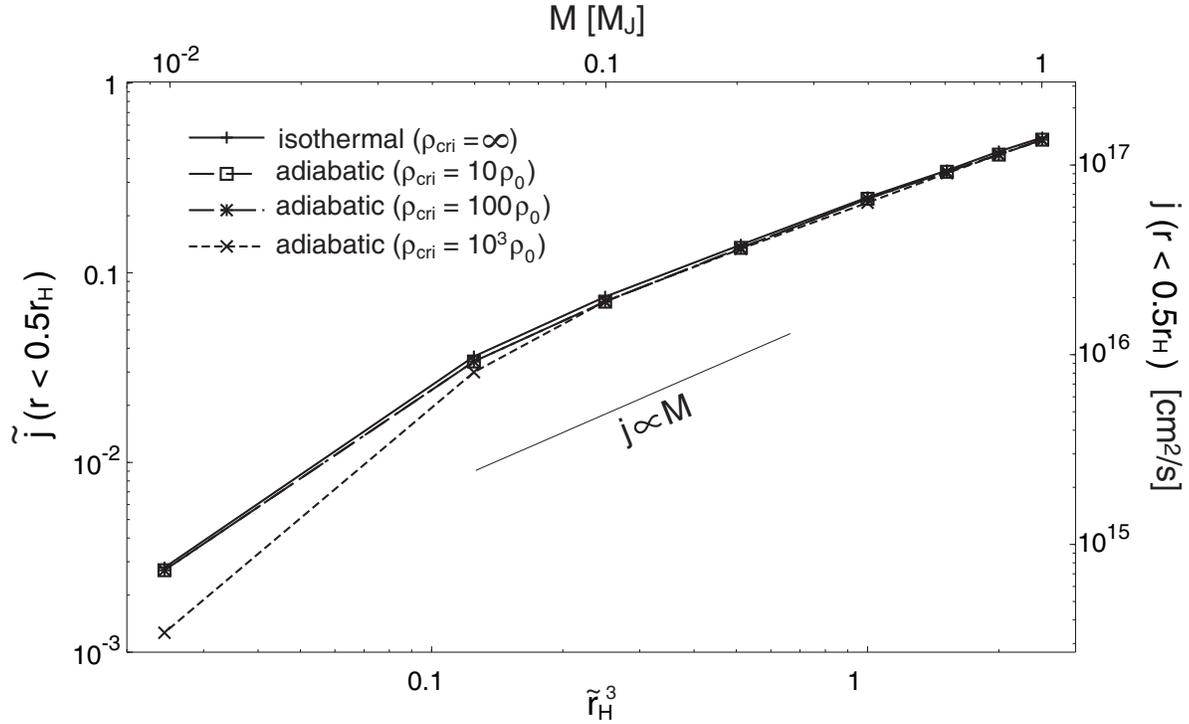}
\caption{
Specific angular momenta for isothermal ($+$) and adiabatic models [$\rho_{\rm cri}=$ 10$\rho_{\rm c,0}$ ($\sq$), 100$\rho_{\rm c,0}$ ($*$), $10^3$$\rho_{\rm c,0}$ ($\times$) ] in the region of  $\tilde{r} < 0.5\, \rht$ against $\rht^3$, which corresponds to the protoplanet's mass [see eq.~(\ref{eq:mass-to-hill})]. 
The upper and right axes indicate the dimensional mass in units of the Jovian mass and specific angular momentum at the Jovian orbit, respectively.
}
\label{fig:5}
\end{center}
\end{figure}

\clearpage
\begin{figure}
\begin{center}
\includegraphics[width=160mm]{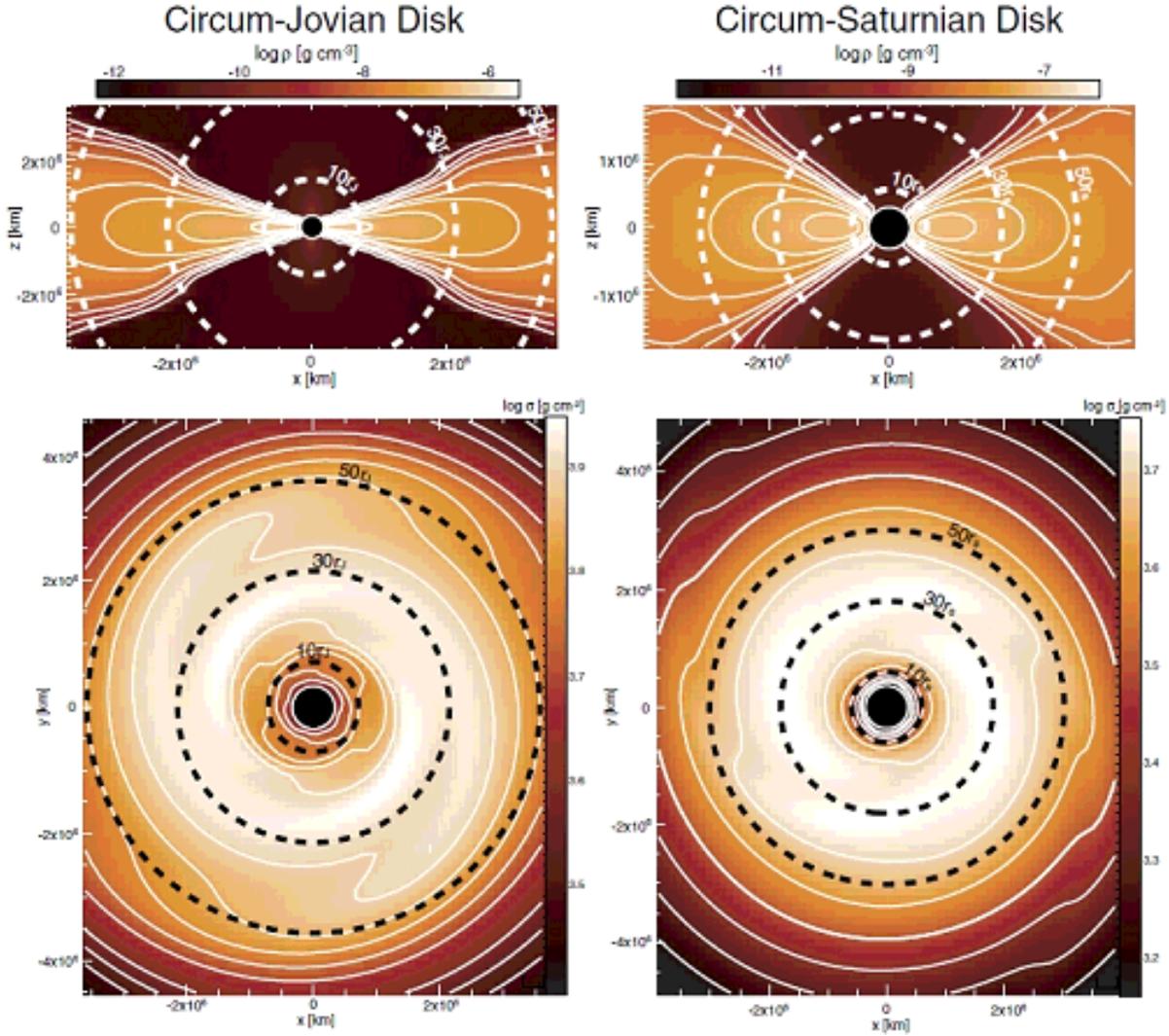}
\caption{
Circum-Jovian (left: model M1A2) and circum-Saturnian (right: model M02A2) disks around the protoplanet.
The density distribution on the $y=0$ plane (upper) and surface density along the $z$-axis (lower) are plotted.
Dimensional units are at Jovian (5.2AU; left) and Saturnian (9.6\,AU; right) orbits.
The dotted lines represent 10, 30 and 50 times the protoplanet's radius, in which present Jovian (left panels) and Saturnian (right panels) radii are adopted. 
}
\label{fig:6}
\end{center}
\end{figure}

\clearpage
\begin{figure}
\begin{center}
\includegraphics[width=160mm]{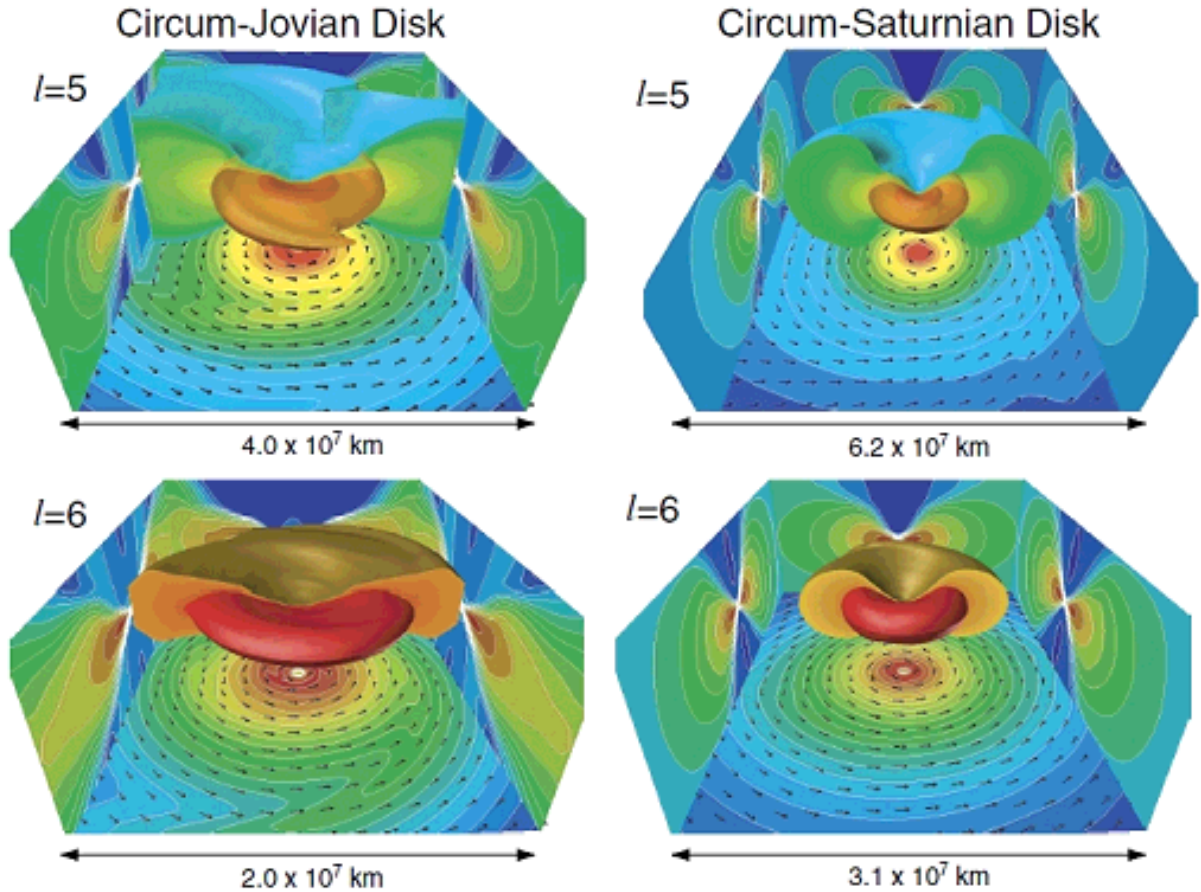}
\caption{
Structures around proto-Jupiter (left: model M1A2) and proto-Saturn (right: model M02A2) in a bird's-eye view.
Density distributions on $x=0$, $y=0$, and $z=0$ planes are plotted on each wall surface.
The colour surfaces indicate iso-density surfaces: $\rho=10^3\rho_{\rm c,0}$ (red), $\rho=10^2\rho_{\rm c,0}$ (orange), and $\rho=10\rho_{\rm c,0}$ (green and blue).
Velocity vectors ({\it arrows}) are plotted on the bottom wall.
Dimensional units are at Jovian (5.2AU; left) and Saturnian (9.6\,AU; right) orbits.
The size of the domain is shown in each panel.
}
\label{fig:7}
\end{center}
\end{figure}

\end{document}